\documentclass[letterpaper, 10 pt, conference]{ieeeconf}
\usepackage{mathtools, cuted}
\usepackage{amssymb}
\usepackage{balance}
\usepackage{mathrsfs}
\usepackage{comment}
\usepackage{mathdots}
\usepackage{xfrac}
\usepackage{bm}
\usepackage{array}
\usepackage{color}
\usepackage{algorithm}
\usepackage{algorithmicx}
\usepackage{algpseudocode}
\usepackage{physics}
\usepackage{comment}
\usepackage{lipsum}
\usepackage{float}
\usepackage{bbm}

\newtheorem{problem}{Problem}

\DeclareMathAlphabet\mathbfcal{OMS}{cmsy}{b}{n}

\usepackage{etoolbox}

\providecommand{\keywords}[1]{\textbf{\textit{Index terms---}}}

\IEEEoverridecommandlockouts                              
\title{\LARGE \bf
Filtering in Multivariate Systems with Quantized Measurements using a Gaussian Mixture-Based Indicator Approximation
}

\author{Angel L. Cedeño, Rodrigo A. Gonz\'alez, Boris I. Godoy and Juan C. Agüero 
\thanks{This work was supported in part by the grants ANID-Fondecyt 3240181 and 1211630, by the ANID-Basal Project AFB240002 (AC3E), and by E2TECH, University of Santiago of Chile. A. L. Cede\~no is with the Faculty of Engineering, Electrical Engineering Department, University of Santiago of Chile, Santiago, Chile. Juan C. Agüero is with the Electronic Engineering Department of Universidad T\'ecnica Federico Santa Mar\'ia, Valpara\'iso, Chile. R. A. Gonz\'alez is with the Department of Mechanical Engineering of Eindhoven University of Technology, Eindhoven, The Netherlands. Boris I. Godoy is with the Department of Mechanical Engineering, Boston University, Boston, USA. Email of the corresponding author: angel.cedeno@usach.cl.}}

\renewcommand{\baselinestretch}{0.95}

\begin{document}

\maketitle
\thispagestyle{empty}
\pagestyle{empty}
\begin{abstract}
This work addresses the problem of state estimation in multivariable dynamic systems with quantized outputs, a common scenario in applications involving low-resolution sensors or communication constraints. A novel method is proposed to explicitly construct the probability mass function associated with the quantized measurements by approximating the indicator function of each region defined by the quantizer using Gaussian mixture models. Unlike previous approaches, this technique generalizes to any number of quantized outputs without requiring case-specific numerical solutions, making it a scalable and efficient solution. Simulation results demonstrate that the proposed filter achieves high accuracy in state estimation, both in terms of fidelity of the filtering distributions and mean squared error, while maintaining significantly reduced computational cost.
\end{abstract}

\section{Introduction}
State estimation is a fundamental problem in the control and monitoring of dynamic systems whose goal is to reconstruct the internal variables (states) of a system from partial and noisy measurements. This field, which studies methods for inferring states from input-output measurements \cite{anderson1979}, is essential in applications such as system identification \cite{Cedeno2024Id,schon2011}, industrial control \cite{Liu2021}, navigation systems \cite{Chiang2012}, sensor networks \cite{Schizas2008}, fault detection \cite{Huang2021}, among others. Within the Bayesian framework, recursive filters update the posterior probability density function (PDF) of the state based on observations, with the Kalman filter and its extensions to handle nonlinear systems being particularly prominent. When measurements exhibit severe nonlinearities such as quantization, the design of efficient estimators becomes a significant challenge, requiring specialized approaches to mitigate bias and information loss.

Quantization, which occurs when measurements are represented using a discrete set of values, typically arises due to limitations in low-cost sensors or communication constraints. This nonlinearity introduces significant distortions in the observations, leading to estimation errors when not properly taken into account in the filtering algorithms \cite{widrow2008quantization}. In particular in critical applications such as navigation systems or networked control systems with limited-capacity channels, ignoring quantization effects can result in biased estimates, instability, or even catastrophic failures. Furthermore, the non-differentiable nature of quantization complicates the use of traditional methods based on linearization, requiring more sophisticated approaches to ensure robust performance.

In response to these challenges, various strategies have been developed to address state estimation in systems with quantized outputs. Among them, particle-based methods stand out, capturing the nonlinearity through stochastic sampling, as well as methods that explicitly model quantization as a probabilistic function within the Bayesian framework. One approach is to approximate the quantizer using gradient-based methods \cite{wigren1995approximate}, and another is to model the quantizer as additive uniform noise \cite{widrow2008quantization,Gustafsson2009}. In \cite{Cedeno2021}, the probability mass function (PMF) of quantized outputs is modeled through an integral equation, which is explicitly approximated using a Gauss-Legendre quadrature rule. This idea allows for the approximation of the likelihood function using a Gaussian mixture model (GMM) structure, which, when propagated through the recursive Bayesian filtering equations, results in GMM-based filtering and prediction PDFs. In addition, optimization techniques and adaptive filters have been proposed to incorporate quantization information and improve estimation accuracy. However, despite these efforts, challenges still remain in balancing computational complexity and estimation accuracy, motivating the search for new algorithms that combine efficiency and robustness in practical settings.

In this work, building on the approach in \cite{Cedeno2021b}, we propose a more efficient method to solve the integral equation that defines the PMF of quantized outputs. This integral, which represents a Gaussian integral over a hyperrectangle defined by the quantizer, is reformulated as an indefinite integral over the entire output dimension space by incorporating an indicator function. Then, recognizing that the indicator function in the interval $[0,1]$ can be interpreted as a uniform distribution over that same interval, we train offline a GMM model to approximate that uniform distribution with the desired accuracy. Moreover, for handling an arbitrary $p$-dimensional hyperrectangle, we construct an approximation by taking the $p$-fold tensor product of the trained GMM, and scale its weights, means, and covariances accordingly. This allows us to explicitly solve the integral equation that defines the PMF, which produces a likelihood function in the form of a Gaussian mixture which is used to construct an explicit state estimator for the system based on the Gaussian sum filter \cite{kitagawa1994two}. Compared to \cite{Cedeno2021b}, which is limited to a single quantized output, our approach generalizes to any output dimension $p$. Once a GMM is trained to approximate the uniform distribution over $[0,1]$, it can be directly extended to higher dimensions, making the method highly scalable.

The remainder of the paper is as follows. In Section \ref{sec:problemformulation} the setup and problem formulation are stated. In Section \ref{sec:bayesian} the approximation of the indicator function is proposed, which allows the Bayesian filtering method outlined in Section \ref{sec:gsf}. Extensive numerical simulations are presented in Section \ref{sec:simulations}, and final remarks are given in Section \ref{sec:conclusions}.

\section{Setup and problem formulation} \label{sec:problemformulation}
\label{sec:definicion}
\subsection{System setup}
\begin{figure}
	\centering
	\includegraphics[width=0.65\linewidth]{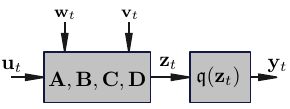}
        \vspace{-0.2cm}
	\caption{Block diagram of a MIMO state-space model with quantized output.}
        \vspace{-0.3cm}
	\label{fig:ssdiagram_nonlinear_functions}
\end{figure}

\begin{figure}
	\centering
	\includegraphics[width=0.8\linewidth]{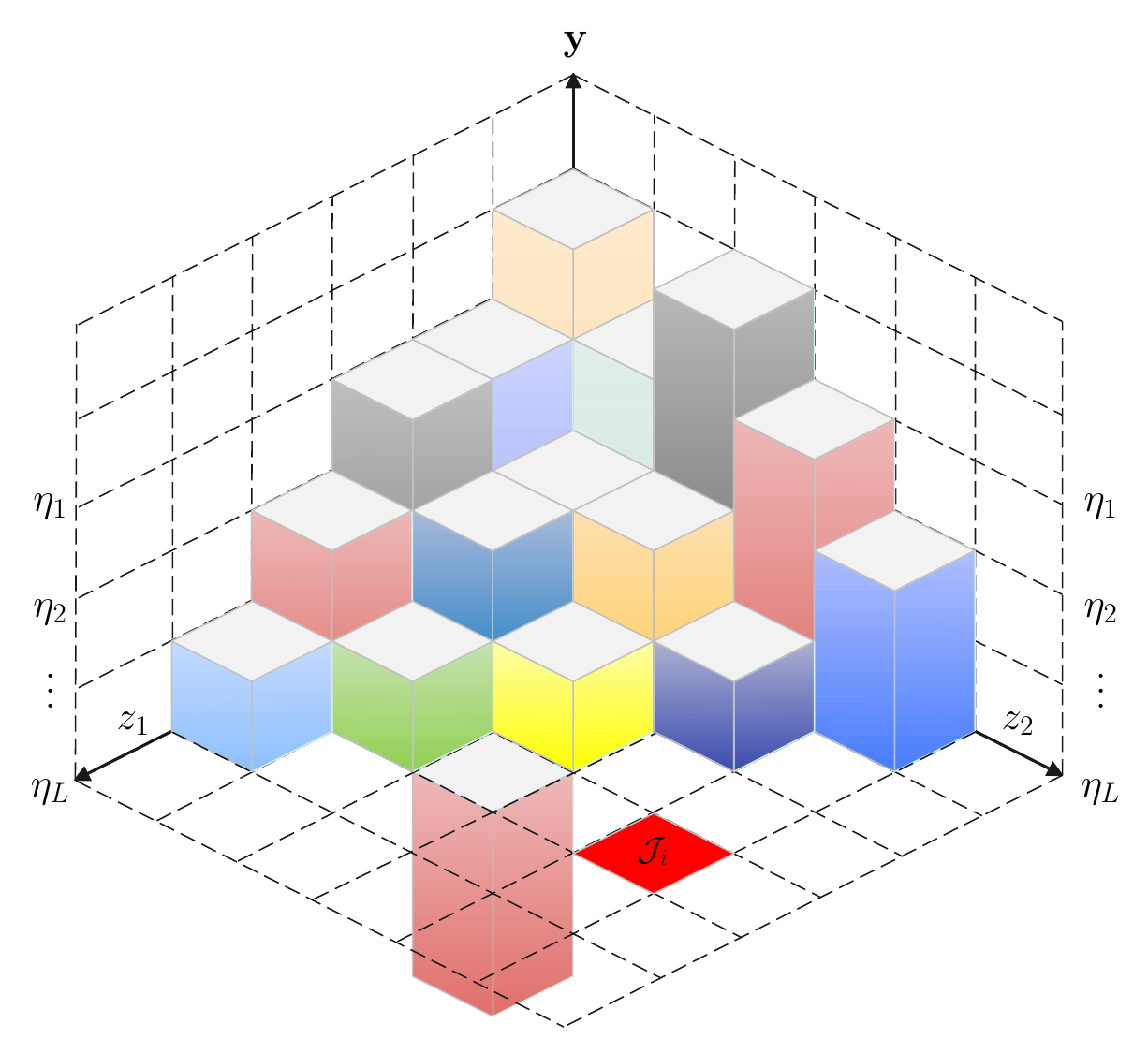}
    \vspace{-0.3cm}
	\caption{Example of multivariate quantizer: case for $p=2$.}
            \vspace{-0.3cm}
	\label{fig:bivariate_quantizer_diagram}
\end{figure}

Consider the following linear time-invariant, multi-input multi-output (MIMO), discrete-time model in state-space representation with quantized outputs (see Fig. \ref{fig:ssdiagram_nonlinear_functions}), given by 
\begin{align}
	\mathbf{x}_{t+1}&=\mathbf{A}\mathbf{x}_{t}+\mathbf{B}\mathbf{u}_{t}+\mathbf{w}_{t}, \label{eqn:general_system_state}\\
	\mathbf{z}_{t}&=\mathbf{C}\mathbf{x}_{t}+\mathbf{D}\mathbf{u}_{t}+\mathbf{v}_{t},\label{eqn:general_system_output_lin}\\
	\mathbf{y}_{t}&=\mathfrak{q}(\mathbf{z}_{t}),\label{eqn:general_system_output}
\end{align}
where $\mathbf{x}_{t} \in \mathbb{R}^{n}$, $\mathbf{z}_{t} \in \mathbb{R}^{p}$, $\mathbf{y}_{t} \in \mathbb{R}^{p}$, and $\mathbf{u}_{t} \in \mathbb{R}^{m}$ are the state vector, the linear output (non measurable), the quantized output (measurable), and the system input, respectively. The matrices of the system are $\mathbf{A} \in \mathbb{R}^{n\times n}$, $\mathbf{B} \in \mathbb{R}^{n\times m}$, $\mathbf{C} \in \mathbb{R}^{p\times n}$ and $\mathbf{D} \in \mathbb{R}^{p\times m}$. The disturbances present in the state equation, $\mathbf{w}_{t} \in \mathbb{R}^{n}$, and the linear output equation, $\mathbf{v}_{t} \in \mathbb{R}^{p}$, are zero-mean independent Gaussian random variables with covariance matrices $\mathbf{Q}$ and $\mathbf{R}$, respectively. The initial condition $\mathbf{x}_1$ is assumed to be a random variable independent of $\mathbf{w}_{t}$ and $\mathbf{v}_{t}$ distributed as $\mathbf{x}_1\sim\mathcal{N}(\mathbf{x}_1;\bm{\mu}_1,\mathbf{P}_1)$, where $\mathcal{N}(\mathbf{x};\bm{\mu},\mathbf{P})$ represents a Gaussian PDF of mean $\bm{\mu}$ and covariance $\mathbf{P}$ corresponding to the random variable $\mathbf{x}$. The quantizer function $\mathfrak{q}(\cdot)$, as seen in Fig. \ref{fig:bivariate_quantizer_diagram}, is given by:
\begin{align}
    \mathfrak{q}(\mathbf{z}_{t}) = 
    \begin{cases}
    	\bm{\eta}_1 & \text{if } \mathbf{z}_t \in \mathcal{J}_1, \\
    	\bm{\eta}_2 & \text{if } \mathbf{z}_t \in \mathcal{J}_2, \\
    	\vdots \\
    	\bm{\eta}_L & \text{if } \mathbf{z}_t \in \mathcal{J}_L, \\
        \vdots
    \end{cases}
\end{align}
Here, $\mathbf{z}_t$ is the input vector at time $t$, and $\bm{\eta}_i$ is the constant value (e.g., centroid or representative value) associated with the region $\mathcal{J}_i$. The quantizer divides the space $\mathbb{R}^p$ into disjoint regions $\mathcal{J}_i$, where each region $\mathcal{J}_i$ is a hyperrectangle defined by lower and upper bounds along each dimension. Formally, each region $\mathcal{J}_i$ is defined as:
\begin{align}\label{eqn:regionsJi}
	\mathcal{J}_i = \{ \mathbf{z} \in \mathbb{R}^p : q_{i-1,j} \leq z_j < q_{i,j}, \forall j = 1, 2, \dots, p \},
\end{align}
 where $\mathbf{z} = [z_1, z_2, \dots, z_p]^\top$ is a vector in $ \mathbb{R}^p$, and $q_{i-1,j}$ with $q_{i,j}$ are real numbers that describe the lower and upper bounds of the $j$th dimension of the region $\mathcal{J}_i$. The bounds for each dimension of $\mathcal{J}_i$ are assumed to be uniformly spaced such that $q_{i,j} = q_{i-1,j} + \Delta_j$, where the quantization step size $\Delta_j$ is the width of each interval along the $j$th dimension. The regions $\mathcal{J}_i$ form a partition of $\mathbb{R}^p$, which means they are disjoint and cover the entire space. An example of this quantization scheme in $\mathbb{R}^2$ is illustrated in Fig. 2.

\subsection{Problem formulation}
The problem addressed in this paper is described below.
\begin{problem}
	\label{prob1}
	Given the set of measurements $\mathbf{y}_{1:t}=\{ \mathbf{y}_1,\mathbf{y}_2$, $\dots$, $\mathbf{y}_t\}$ of the nonlinear output and the set of input data $\mathbf{u}_{1:t}=\left\lbrace \mathbf{u}_1,\mathbf{u}_2,\dots,\mathbf{u}_t\right\rbrace$, for each time instant $t\in \mathbb{N}$,
	\begin{enumerate}
		\item Obtain the conditional PDF of the state vector $\mathbf{x}_t$, $p(\mathbf{x}_t|\mathbf{y}_{1:t})$, and
		\item Derive a state estimator and the covariance matrix of the estimation error:
		\begin{align}
			\hat{\mathbf{x}}_{t|t}&=\mathbb{E}\left\lbrace \mathbf{x}_t|\mathbf{y}_{1:t}\right\rbrace, \label{eqn:filtered_state} \\
			\bm{\Sigma}_{t|t}&=\mathbb{E}\left\lbrace (\mathbf{x}_t-\hat{\mathbf{x}}_{t|t})(\mathbf{x}_t-\hat{\mathbf{x}}_{t|t})^{\top}|\mathbf{y}_{1:t}\right\rbrace, \label{eqn:filtered_cov_state}
		\end{align}
		where $\mathbb{E}\left\lbrace \mathbf{x}|\mathbf{y} \right\rbrace$ is the conditional expectation of the random variable $\mathbf{x}$ given the variable $\mathbf{y}$.
	\end{enumerate} 
\end{problem}

\section{Bayesian framework for filtering under quantization}
\label{sec:bayesian}
The Bayesian approach for state estimation is a probabilistic framework used to infer the state of a dynamic system given a sequence of noisy observations. The goal is to recursively update the belief about the state $\mathbf{x}_{t} $ at time $t$ based on the prior belief and the new observation $ \mathbf{y}_{t} $. This is achieved through two key PDFs: the prior PDF $ p(\mathbf{x}_{t} | \mathbf{y}_{1:t-1})$, which represents the belief about the state before incorporating the new observation, and the posterior PDF $ p(\mathbf{x}_{t} | \mathbf{y}_{1:t}) $, which updates this belief after considering the current measurement $\mathbf{y}_t$. The process is governed by two recursive equations: the prediction step, which propagates the state forward using the system model $ p(\mathbf{x}_{t+1} | \mathbf{x}_{t}) $, and the update step, which refines the estimate using the likelihood $ p(\mathbf{y}_{t} | \mathbf{x}_{t}) $ derived from the observation model. These steps are expressed as:
\begin{align}
    p(\mathbf{x}_{t} | \mathbf{y}_{1:t}) &\propto p(\mathbf{y}_{t} | \mathbf{x}_{t}) \, p(\mathbf{x}_{t} | \mathbf{y}_{1:t-1}), \label{eqn:bayes_measurement}\\
	p(\mathbf{x}_{t+1} | \mathbf{y}_{1:t}) &= \int_{\mathbb{R}^n} p(\mathbf{x}_{t+1} | \mathbf{x}_{t}) \, p(\mathbf{x}_{t} | \mathbf{y}_{1:t}) \text{d}\mathbf{x}_{t}. \label{eqn:bayes_time}
\end{align}
The prior PDF captures the dynamical behavior of the system, while the likelihood function $p(\mathbf{y}_{t} | \mathbf{x}_{t})$ relates the state to the observations. The posterior PDF provides the optimal estimate under the Bayesian framework. However, exact solutions are often intractable for nonlinear systems, leading to the use of approximations such as the extended Kalman filter or particle filter \cite{sarkka2013bayesian}. 

Following the approach in \cite{Cedeno2021b}, we can write the distribution $p(\mathbf{y}_{t}|\mathbf{x}_{t})$ as:
\begin{align}
    p(\mathbf{y}_{t}|\mathbf{x}_{t}) = \int_{\{\mathbf{z}_t\colon \mathbf{z}_t\in \mathcal{J}_{i}\}} \hspace{-0.2cm} \mathcal{N}(\mathbf{z}_t;\mathbf{C}\mathbf{x}_{t}+\mathbf{D}\mathbf{u}_{t},\mathbf{R})\text{d}\mathbf{z}_t,
\end{align}
where the sets $\mathcal{J}_{i}$ correspond to those defined in \eqref{eqn:regionsJi}. Unlike \cite{Cedeno2021b}, which addresses a multi-input single-output (MISO) system, this work considers a MIMO system, where the multivariable output is determined by a quantization function that defines a fixed $p$-dimensional hyperplane for each set $\mathcal{J}_{i}$. This represents a much more challenging and general problem than the one addressed in \cite{Cedeno2021b}, since solving it requires integration over multiple dimensions, which can be computationally prohibitive.

\begin{figure*}
	\centering
	\includegraphics[width=1\linewidth]{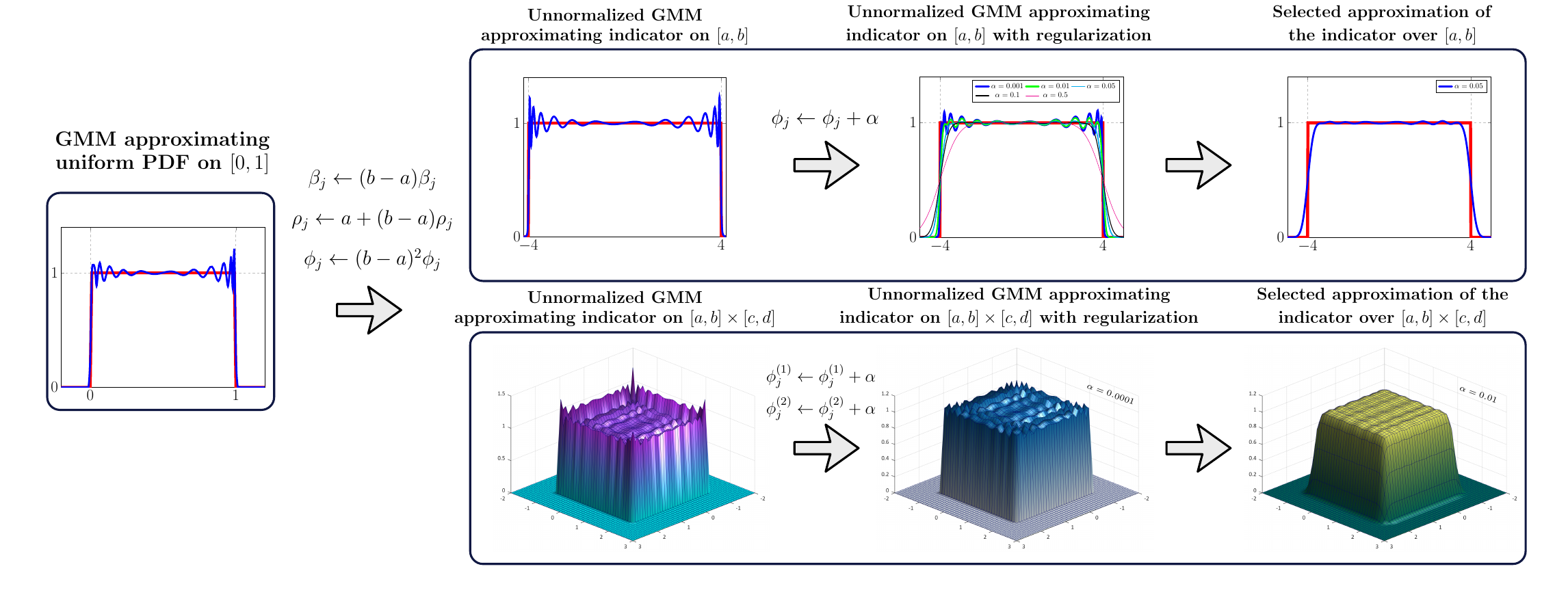}
        \vspace{-0.7cm}
	\caption{Given a GMM that approximates the uniform PDF over the interval $[0,1]$, it is possible to construct approximations of the indicator function on an interval $[a,b]$ by appropriately scaling the weights, means, and covariances of the GMM (see top figures). This yields an unnormalized GMM that approximates the indicator function over the specified interval. Furthermore, by concatenating multiple dimensions, one can obtain a GMM that approximates a $p$-dimensional indicator function (see bottom figures for an example with $p=2$). A regularization term can also be added to each covariance $\phi_j$ to produce a smoother approximation, which can be applied uniformly across all dimensions. }
	\label{fig:indicator_approximation}
    	\vspace{-0.5cm}
\end{figure*}

The solution proposed in this work adopts a novel approach that can also scale easily with the output dimension. First, note that $p(\mathbf{y}_{t}|\mathbf{x}_{t})$ can be rewritten as follows:
\begin{align}
\label{pytxt}
    p(\mathbf{y}_{t}|\mathbf{x}_{t}) \hspace{-0.03cm}=\hspace{-0.07cm} \int_{\mathbb{R}^{p}} \hspace{-0.05cm}\mathcal{I} \hspace{-0.03cm}\left\lbrace \mathbf{z}_t \hspace{-0.03cm}\in\hspace{-0.03cm} \hspace{-0.02cm}\mathcal{J}_{i} \right\rbrace  \mathcal{N}(\mathbf{z}_t;\mathbf{C}\mathbf{x}_{t}\hspace{-0.05cm}+\hspace{-0.05cm}\mathbf{D}\mathbf{u}_{t},\mathbf{R}) \text{d}\mathbf{z}_t,
\end{align}
where $\mathcal{I} \left\lbrace \mathbf{z}_t \in \mathcal{J}_{i} \right\rbrace$ is the indicator function, defined as:
\begin{align}
    \mathcal{I} \left\lbrace \mathbf{z}_t \in \mathcal{J}_{i} \right\rbrace = 
    \begin{cases}
        1, & \textnormal{if } \mathbf{z}_t \in  \mathcal{J}_{i} \\
        0, & \textnormal{if } \mathbf{z}_t \notin  \mathcal{J}_{i}.
    \end{cases}
\end{align}
In order to obtain an explicit form of $p(\mathbf{y}_{t}|\mathbf{x}_{t})$, that is, one that does not depend on an integral, we propose in this work to approximate the indicator function using an unnormalized Gaussian mixture model, as follows:
\begin{align}
\label{gmmapproximation}
    \mathcal{I} \left\lbrace \mathbf{z}_t \in \mathcal{J}_{i} \right\rbrace \approx \sum_{j=1}^{K} \beta_j \mathcal{N}(\mathbf{z}_t;\bm{\rho}_j,\bm{\Phi}_j),
\end{align}
where $\beta_j$, $\bm{\rho}_j$, and $\bm{\Phi}_j$ respectively represent the weights, means, and covariance of the Gaussian components that approximate the indicator function within the region $\mathcal{J}_{i}$. The basic idea is to consider a uniform distribution in the interval $[0,1] \subset \mathbb{R}$ and train offline a GMM to approximate that uniform distribution in the same interval, producing a univariate GMM with parameters $\left\lbrace \beta_j,\rho_j,\phi_j\right\rbrace_{j=1}^{K}$. Note that training the GMM can be carried out by drawing $N_{\textnormal{train}}$ samples from the uniform distribution on $[0,1]$ and applying the Expectation-Maximization algorithm in \cite[Chap. 1]{mengersen2011} to fit a GMM to the sampled data. Then, given the nature of the Gaussian distribution, it is sufficient to scale the weights, means and covariances to approximate the indicator function over any interval $[a,b]$ as follows:
\begin{align}
    \beta_j &\leftarrow  ~(b-a)\beta_j,\label{eqn:scaled_betaj}\\
    \rho_j &\leftarrow ~a + (b - a) \rho_j, \label{eqn:scaled_rhoj}\\
    \phi_j &\leftarrow ~(b - a)^2 \phi_j\label{eqn:scaled_phij},
\end{align}
where the $\beta_j$, $\rho_j$, and $\phi_j$ on the right-hand side of \eqref{eqn:scaled_betaj}, \eqref{eqn:scaled_rhoj} and \eqref{eqn:scaled_phij} represent the parameters of the GMM originally trained over the interval $[0,1]$. Moreover, since the regions considered in this work are hyperrectangles with equal side lengths, a multivariate approximation of the indicator function over $\mathcal{J}_{i}$ can be constructed by taking the tensor product of the univariate GMM with itself $p$ times. This results in a $p$ dimensional Gaussian mixture, where the mean and covariance of each component are formed by concatenating the corresponding univariate parameters, and the weights are obtained as the product of the original weights. Additionally, a positive constant $\alpha$ can be added to the variance as a regularization term to ensure a smoother fit of the model to the uniform density \cite{Wang2024Actuator}, see Figure \ref{fig:indicator_approximation}.

The previously derived approximation of the indicator function leads to the following expression for $p(\mathbf{y}_{t}|\mathbf{x}_{t})$:
\begin{align}
    p(\mathbf{y}_{\hspace{-0.01cm}t}\hspace{-0.01cm}|\mathbf{x}_{t}\hspace{-0.01cm}) \hspace{-0.06cm}&\approx\hspace{-0.1cm} \sum_{j=1}^K\beta_j\hspace{-0.06cm}\int_{\mathbb{R}^{p}} \hspace{-0.15cm} \mathcal{N}(\mathbf{z}_t;\bm{\rho}_j,\bm{\Phi}_j)\mathcal{N}(\mathbf{z}_t;\mathbf{C}\mathbf{x}_{t}\hspace{-0.05cm}+\hspace{-0.05cm}\mathbf{D}\mathbf{u}_{t},\mathbf{R}) \text{d}\mathbf{z}_t \notag \\
    &=\hspace{-0.08cm}\sum_{j=1}^{K} \hspace{-0.05cm}\beta_{\hspace{-0.02cm}j} \mathcal{N}\hspace{-0.02cm}(\bm{\rho}_j\hspace{-0.01cm}; \hspace{-0.03cm}\mathbf{C}\mathbf{x}_{t}\hspace{-0.07cm}+\hspace{-0.07cm}\mathbf{D}\mathbf{u}_{t},\hspace{-0.03cm}\mathbf{R}\hspace{-0.07cm}+\hspace{-0.07cm}\bm{\Phi}_j\hspace{-0.01cm}) \hspace{-0.12cm}\int_{\mathbb{R}^p}\hspace{-0.22cm}\mathcal{N}\hspace{-0.01cm}(\mathbf{z}_t;\hspace{-0.03cm}\mathbf{m},\hspace{-0.03cm}\mathbf{S})\text{d}\mathbf{z}_t \notag \\
    \label{eqn:pytxt}
    &=\sum_{j=1}^{K} \beta_j \, \mathcal{N}(\bm{\rho}_j; \mathbf{C}\mathbf{x}_{t}+\mathbf{D}\mathbf{u}_{t},\mathbf{R}+\bm{\Phi}_j),
\end{align}
where the product of Gaussian densities yields $\mathbf{m}$ and $\mathbf{S}$ that are given in, e.g., \cite[Sec. 8.1.8]{petersen2012}. Note that this particular form of the probability function $p(\mathbf{y}_{t}|\mathbf{x}_{t})$, expressed as a weighted sum of Gaussian distributions, preserves its structure under the propagation defined by the filtering equations in \eqref{eqn:bayes_measurement} and \eqref{eqn:bayes_time}, since the noise terms $\mathbf{w}_t$ and $\mathbf{v}_t$ are Gaussian distributed. This specific form of $p(\mathbf{y}_{t}|\mathbf{x}_{t})$ leads to a filter that, when the Bayesian filtering equations are solved iteratively, corresponds to the traditional Gaussian sum filter (GSF, \cite{alspach1972nonlinear}), with tailored modifications to account for the particular form of the variance in each Gaussian component. 

\section{Gaussian Sum Filtering with quantized measurements}
\label{sec:gsf}
For the system \eqref{eqn:general_system_state} and \eqref{eqn:general_system_output_lin} with quantized measurements given by \eqref{eqn:general_system_output}, and the output model in \eqref{eqn:pytxt}, the Gaussian sum filter is given by the following algorithm. \\
	
\noindent \textbf{Initialization:}
For the time instant $t=1$, the PDF of the prediction step is given by $p(\mathbf{x}_1)=\mathcal{N}(\mathbf{x}_1;\bm{\mu}_1,\mathbf{P}_1)$.\\
 
\noindent \textbf{Correction stage:} For $t=1,\dots,N$, the filtering PDF of the current state $\mathbf{x}_t$ given the quantized output measurements $\mathbf{y}_1,\dots,\mathbf{y}_t$, is the Gaussian mixture model
\begin{equation}\label{eqn:correction} 
    p(\mathbf{x}_t|\mathbf{y}_{1:t})=
    \sum_{\ell=1}^{\mathcal{M}_{t|t}} \delta_{t|t}^{\ell}\mathcal{N}(\mathbf{x}_{t};\hat{\mathbf{x}}_{t|t}^{\ell},\bm{\Gamma}_{t|t}^{\ell}),
\end{equation}
where $\mathcal{M}_{t|t}=K\mathcal{M}_{t|t-1}$, with $K$ being the number of Gaussian components of the approximation of the indicator function. For each pair $(j,k)$, where $j=1,\dots,K$ and $k=1,\dots, \mathcal{M}_{t|t-1}$, we obtain a new index $\ell=(k-1)K+j$ such that the weights, mean values and covariance matrices are given by
\begin{align}
    \delta_{t|t}^{\ell} & \!=\! \bar{\delta}_{t|t}^{\ell}/(\textstyle \sum_{r=1}^{\mathcal{M}_{t|t}}\bar{\delta}_{t|t}^{r}), \label{eqn:lemma_filtering_delta}\\	
    \bar{\delta}_{t|t}^{\ell}&\!=\!\beta_{j}\delta_{t|t-1}^{k}\mathcal{N}(\bm{\rho}_j; \mathbf{C}\hat{\mathbf{x}}_{t|t-1}^{k}\!+\!\mathbf{D}\mathbf{u}_t,\! \mathbf{R}\!+\!\bm{\Phi}_j\!+\!\mathbf{C}\bm{\Gamma}_{t|t-1}^{k}\mathbf{C}^{\top}), \notag \\
    \mathbf{K}_t^k&\!=\!\bm{\Gamma}_{t|t-1}^{k}\mathbf{C}^\top(\mathbf{R}+\bm{\Phi}_j+\mathbf{C}\bm{\Gamma}_{t|t-1}^{k}\mathbf{C}^\top\hspace{-0.01cm})^{-1},\\
    \hat{\mathbf{x}}_{t|t}^{\ell}&\!=\!\hat{\mathbf{x}}_{t|t-1}^{k}+\mathbf{K}_t^k(\bm{\rho}_j-\mathbf{C}\hat{\mathbf{x}}_{t|t-1}^{k}-\mathbf{D}\mathbf{u}_t),\label{eqn:lemma_filtering_x} \\
    \bm{\Gamma}_{t|t}^{\ell}&\!=\!(\mathbf{I}-\mathbf{K}_t^k\mathbf{C})\bm{\Gamma}_{t|t-1}^{k}. \label{eqn:lemma_filtering_gamm}
\end{align}	
The initial values of the recursion are $\mathcal{M}_{1|0}=1$, $\delta_{1|0}=1$, $\hat{\mathbf{x}}_{1|0}=\bm{\mu}_1$, and $\bm{\Gamma}_{1|0}=\mathbf{P}_1$. \\

\noindent \textbf{Prediction stage:} The predicting PDF of the state $\mathbf{x}_{t+1}$,  given the quantized output measurements $\mathbf{y}_1,\dots,\mathbf{y}_t$, is the Gaussian mixture model defined as
\begin{equation}\label{eqn:prediccion}
    p(\mathbf{x}_{t+1}|\mathbf{y}_{1:t})=\sum_{\ell=1}^{\mathcal{M}_{t+1|t}} \delta_{t+1|t}^{\ell}\mathcal{N}(\mathbf{x}_{t+1};\hat{\mathbf{x}}_{t+1|t}^{\ell},\bm{\Gamma}_{t+1|t}^{\ell}),
\end{equation}
where $\mathcal{M}_{t+1|t}=\mathcal{M}_{t|t}$, and the weights, mean values, and covariance matrices are given by
\begin{align}
    \delta_{t+1|t}^{\ell}&=\delta_{t|t}^{\ell}, \label{eqn:lemma_filtering_time_delta}\\
    \hat{\mathbf{x}}_{t+1|t}^{\ell}&=\mathbf{A}\hat{\mathbf{x}}_{t|t}^{\ell}+\mathbf{B}\mathbf{u}_t, \label{eqn:lemma_filtering_time_x}\\	
    \bm{\Gamma}_{t+1|t}^{\ell}&=\mathbf{Q}+\mathbf{A}\bm{\Gamma}_{t|t}^{\ell}\mathbf{A}^\top. \label{eqn:lemma_filtering_time_gamm}
\end{align}
Given the PDFs $p(\mathbf{x}_t | \mathbf{y}_{1:t})$ modeled as GMMs, the state estimator $\hat{\mathbf{x}}_{t|t}$ and the estimation error covariance $\bm{\Sigma}_{t|t}$ can be easily computed using the standard expectation rules \cite{Cedeno2021b}.

During the correction stage, the number of components grows iteratively according to $\mathcal{M}_{t|t} = K\mathcal{M}_{t|t-1}$. For example, if initially $\mathcal{M}_{1|0} = 10$ and $K = 10$, the number of components would increase according to $\mathcal{M}_{t|t} = 10^{t+1}$, highlighting the need for reduction methods to control computational complexity. Gaussian mixture reduction techniques, such as K-means clustering, the Gaussian merging algorithm, and the Integral Square Difference (ISD) method, aim to approximate a Gaussian mixture with a smaller number of components while preserving the accuracy of the original distribution, see e.g., \cite{kitagawa1994two}. K-means clustering groups nearby Gaussians into clusters, offering computational efficiency but potentially losing fine details. The merging algorithm combines similar components based on metrics such as the Kullback–Leibler (KL) divergence, better preserving the structure of the distribution at the cost of increased computational complexity. Finally, the ISD-based method enables a structured reduction of Gaussian mixtures, outperforming previous approaches when dealing with a large number of components, at the cost of a higher computational burden. In this work, the KL divergence approach for Gaussian sum reduction proposed in \cite{runnalls2007kullback} is adopted. 

\section{Simulations}
\label{sec:simulations} 
In this section we compare the proposed method, referred to as GSF, with three well-established approaches from the literature: the particle filter (PF) in its bootstrap version \cite{gordon1993novel,doucet2000sequential} with 300 particles, the unscented Kalman Filter (UKF) \cite{Julier1997}, and the quantized Kalman filter (QKF) \cite{gomez2020}. To obtain a reference PDF for comparison, we use a particle filter with 20,000 particles. These reference PDFs are referred to as the ground truth (GT). First, a SISO scalar system is implemented to visualize the estimated filtering PDFs, which becomes more challenging in higher-dimensional systems. Afterwards, a MIMO system is considered in Example 2. For all examples, box-plot graphs of the estimation error are provided, comparing the true state and the estimated state. The simulations were performed on a computer running Windows 11, with MATLAB R2024b, equipped with an AMD Ryzen 9 5900HX with a Radeon Graphics processor at 3.30 GHz and 32 GB of RAM.

\subsection{Example 1} 
Consider the following state-space system defined by the matrices: $A = 0.8$, $B = 1.5$, $C = 2.8$, and $D = 1.8$. The quantization function is given by $y_t = \Delta \textnormal{round}(z_t/\Delta)$, where $\Delta=10$. The process and measurement noise are distributed as $w_t \sim \mathcal{N}(w_t;0,1)$ and $v_t \sim \mathcal{N}(v_t;0,0.1)$, respectively. The input signal $u_t$ is sampled from $\mathcal{N}(u_t;0,2)$, and the initial condition is given by $x_1 \sim \mathcal{N}(x_1;1,2)$. The indicator function for the interval $[0,1]$ was trained offline using the \texttt{fitdist} function in MATLAB with 20 Gaussian components and $10^{7}$ samples drawn from a uniform distribution over $[0,1]$. 
\begin{figure}
	\centering
	\includegraphics[width=1\linewidth]{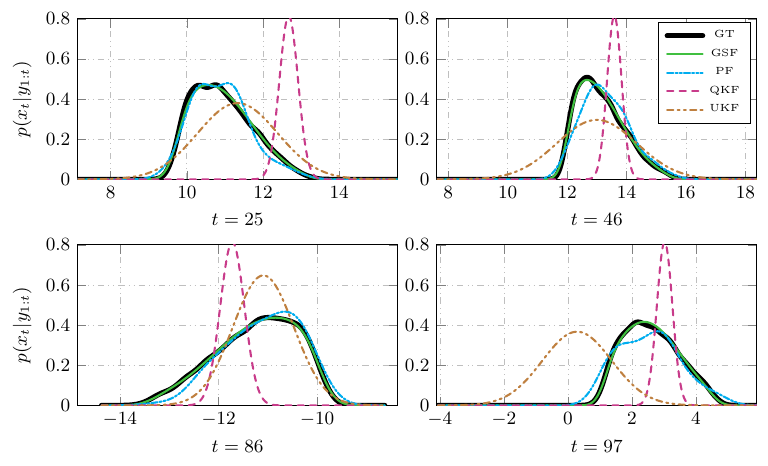}
    \vspace{-0.8cm}
	\caption{Example 1: Filtering PDFs $p(x_t | y_{1:t})$ at selected time steps.}
    \label{fig:pdf_filtering_1states}
\end{figure}
\begin{figure}
	\centering
	\includegraphics[width=1\linewidth]{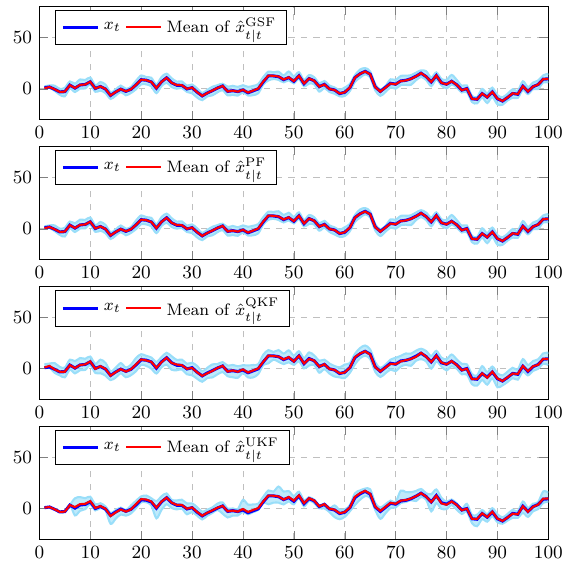}
    \vspace{-0.7cm}
	\caption{Example 1: Monte Carlo simulation with 100 runs. The light-blue shaded region contains all estimated state sequences across different realizations of the noises $w_t$ and $v_t$. The red line represents the mean of all estimates, while the blue line corresponds to the true state sequence.}
    \vspace{-0.7cm}
    \label{fig:cloud_filtering_1states}
\end{figure}
\begin{figure}
	\centering
	\includegraphics[width=1\linewidth]{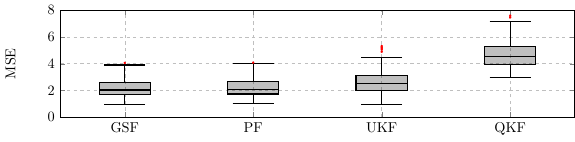}
	\vspace{-0.7cm}
    \caption{Example 1: MSE between estimated $\hat{x}_{t|t}$ and true $x_t$states.}
    \label{fig:state_boxplot_mse_example1}
\end{figure}
In Fig.~\ref{fig:pdf_filtering_1states}, the filtering PDFs at selected time steps are shown. It can be observed that the proposed method, GSF, closely matches the ground truth PDFs (GT), followed by the PF filter. However, the other methods produce PDFs that significantly deviate from GT. Likewise, Fig.~\ref{fig:cloud_filtering_1states} shows the estimated state sequences from a Monte Carlo simulation with 100 realizations. In terms of state estimates (i.e., the means of the corresponding distributions), the GSF and PF filters provide the most accurate estimates, followed by the UKF and, finally, the QKF. Additionally, Fig.~\ref{fig:state_boxplot_mse_example1} presents the Mean Squared Errors (MSE) between the estimated state sequence $\hat{x}_{t|t}$ and the true state sequence $x_t$. As seen in this figure, the same performance trend observed in the previous figure is maintained: the GSF and PF filters yield the most accurate estimates, followed by the UKF and then the QKF. From these results, it can be concluded that the proposed method produces the best estimates of the state PDFs and their corresponding means, followed by the particle filter—whose accuracy can be improved by increasing the number of particles, albeit at a higher computational cost—then the UKF, and finally the QKF.
\subsection{Example 2}
\begin{figure}
	\centering
	\includegraphics[width=1\linewidth]{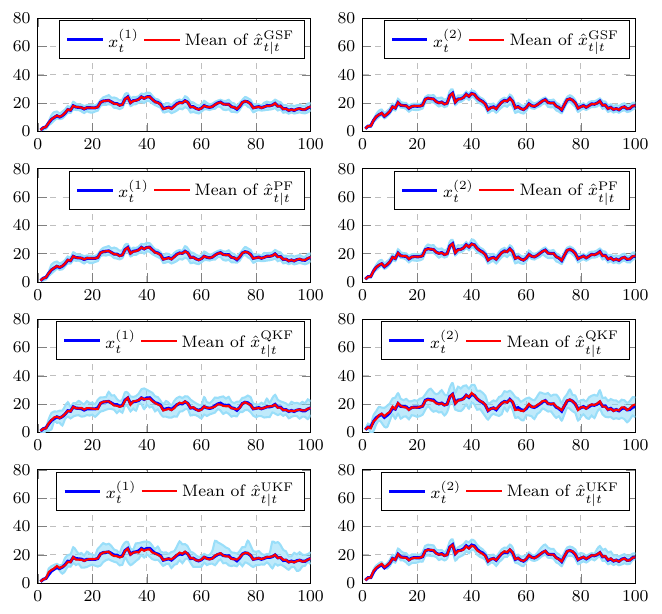}
	    \vspace{-0.7cm}
    \caption{Example 2: Monte Carlo simulation with 100 runs. The light-blue shaded region contains all estimated state sequences across different realizations of the noises $\mathbf{w}_t$ and $\mathbf{v}_t$. The red line represents the mean of all estimates, while the blue line corresponds to the true state sequence.}
    \label{fig:cloud_filtering_2states}
    \vspace{-0.4cm}
\end{figure}

\begin{figure}
	\centering
	\includegraphics[width=0.97\linewidth]{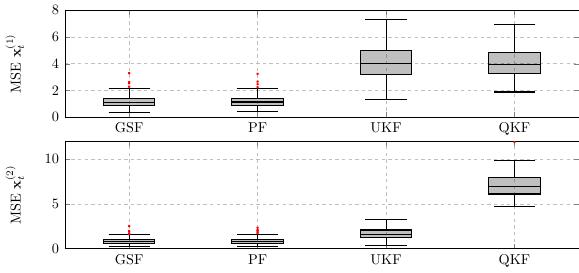}
	    \vspace{-0.35cm}
    \caption{Example 2: Mean Squared Error (MSE) between the estimated state sequence $\hat{\mathbf{x}}_{t|t}$ and the true state sequence $\mathbf{x}_t$.}
    \vspace{-0.7cm}
    \label{fig:state_boxplot_mse_example2}
\end{figure}

Consider the following state-space system defined by
\begin{align}
	\mathbf{x}_{t+1}&=
    \begin{bmatrix}
        0.7362&0.1636\\0.1636&0.7362
    \end{bmatrix}\mathbf{x}_{t}+
    \begin{bmatrix}
        0.8&0.4\\1.2&0.4
    \end{bmatrix}\mathbf{u}_{t}+\mathbf{w}_{t},\\
	\mathbf{z}_{t}&=
    \begin{bmatrix}
        1.05&0.35\\1.40&0.70
    \end{bmatrix} \mathbf{x}_{t}+
    \begin{bmatrix}
        0.40&0.80\\1.20&0.16
    \end{bmatrix}\mathbf{u}_{t}+\mathbf{v}_{t},
\end{align}
with the quantization function being given by $\mathbf{y}_{t} = \Delta \textnormal{round}(\mathbf{z}_t/\Delta)$, where $\Delta=7$. The process and measurement noises follow the distributions $\mathbf{w}_t \sim \mathcal{N}(\mathbf{w}_t; \mathbf{0}, 0.5\mathbf{I}_2)$ and $\mathbf{v}_t \sim \mathcal{N}(\mathbf{v}_t; \mathbf{0}, 0.1\mathbf{I}_2)$, respectively. The input $\mathbf{u}_t$ is drawn from a normal distribution $\mathcal{N}(\mathbf{u}_t; [1,2]^{\top}, 2\mathbf{I}_2 )$, and the initial state is assumed to follow $\mathbf{x}_1 \sim \mathcal{N}(\mathbf{x}_1; [1,2]^{\top}, 0.01\mathbf{I}_2)$, where $\mathbf{I}_2$ denotes the identity matrix of order 2.
Figure~\ref{fig:cloud_filtering_2states} shows the estimated state trajectories obtained from a Monte Carlo simulation with 100 realizations. From this figure, it is evident that, in terms of state estimation accuracy, the GSF and PF filters perform best, followed by the UKF, with the QKF showing the least accurate results. Figure~\ref{fig:state_boxplot_mse_example2} presents the Mean Squared Error (MSE) between the estimated state sequence $\hat{\mathbf{x}}_{t|t}$ and the true sequence $\mathbf{x}_t$, where $\mathbf{x}_t^{(1)}$ and $\mathbf{x}_t^{(2)}$ denote the first and second state variables, respectively. The results confirm the same performance trend observed previously: GSF and PF provide the most accurate estimates, followed by UKF and finally QKF. Table~\ref{tabla:tiempos} shows the computation times (in the format mean \( \pm \) standard deviation) for each filter across the 100 realizations of the Monte Carlo experiment. As observed, the Kalman-based filters—UKF and QKF—exhibit the lowest computation times, followed by the GSF filter, which shows only a slight increase in runtime as the number of outputs grows, despite also performing a Gaussian mixture reduction step. On the other hand, the particle filter, using the same number of particles, experiences a drastic increase in computation time—approximately 80-fold—when the number of outputs increases from 1 to 2. 

\begin{table}[]
\caption{Example 2: Computation time (in seconds) of each filter.}
\resizebox{\columnwidth}{!}{
\begin{tabular}{|c|c|c|c|c|}
\hline
     Example     & GSF & PF & UKF & QKF \\ \hline
 1 & $0.3214  \pm 0.0160$ & $0.0328  \pm 0.0048$   & $0.0112  \pm 0.0042$    &  $0.0017  \pm 0.0005$   \\ \hline
 2 & $0.7582  \pm  0.0488$    &  $2.6282  \pm  0.1161$  & $0.0187  \pm  0.0049$    &  $0.0022  \pm  0.0007$   \\ \hline
\end{tabular}}
\label{tabla:tiempos}
\end{table}

\section{Conclusions}
\label{sec:conclusions}
	    \vspace{-0.1cm}
This work has addressed the problem of state estimation in multivariable dynamic systems with quantized outputs by proposing a novel approach based on the explicit construction of the PMF of the quantized output. This is achieved by approximating the indicator function of each region defined by the quantizer using GMMs. Unlike previous methods that solved the integral defining the PMF using Gaussian quadrature, the proposed solution is more practical and efficient: it consists of approximating the uniform distribution over the interval $[0,1]$ with a GMM model of arbitrary precision in an offline fashion. This approximation can be extended to systems with multiple quantized outputs via the tensor product of the one-dimensional model. This innovative strategy enables an explicit representation of the PMF as a Gaussian mixture, facilitating the implementation of a robust and accurate recursive Bayesian estimator, even in the presence of severe nonlinearities induced by quantization. The numerical results show that the proposed method outperforms traditional filters such as the particle filter, the unscented Kalman filter, and the quantized Kalman filter. In the conducted experiments, the method achieves a closer approximation to the true distributions (ground truth) and lower mean squared errors, performing exceptionally well in scenarios where quantization introduces significant distortions. Furthermore, the results show that for multivariate systems, the proposed method has a relatively low computational cost, comparable to Kalman-based filters, and is faster than a particle filter with a number of particles that achieves similar accuracy. Future work includes considering saturated quantizers and extending this approach to Wiener models.

\bibliographystyle{plain}
\bibliography{bibliography.bib}

\end{document}